\documentclass[letter]{aa} 
\usepackage{graphicx}
\usepackage{txfonts}
\newcommand{\uc}{UC H{\sc ii}}
\newcommand{\kms}{km~s$^{-1}$}
\newcommand{\hco}{HCO$^{+}$(1--0)}
\newcommand{\lm}{$L_\mathrm{bol}/M$}

\begin{document}
   \title{SiO outflows in high-mass star forming regions: A potential chemical clock?\thanks{Based on observations conducted with the IRAM~30-m telescope near Pico Veleta (Granada, Spain), and the Atacama Pathfinder Experiment (APEX) ESO project: 181.C-0885.. IRAM is supported by INSU/CNRS (France), MPG (Germany), and IGN (Spain). APEX is a collaboration between the Max-Planck-Institut f\"ur Radioastronomie, the European Southern Observatory, and the Onsala Space Observatory.}}

   \subtitle{}

   \author{A. L\'opez-Sepulcre
          \inst{1}
          \and
          C.M. Walmsley
          \inst{1}
          \and
          R. Cesaroni
          \inst{1}
          \and
          C. Codella
          \inst{1}
          \and
          F. Schuller
          \inst{2}
          L. Bronfman
          \inst{3}
          \and 
          S.J. Carey
          \inst{4}
          \and 
          K. M. Menten
          \inst{2}
          \and 
          S. Molinari
          \inst{5}
          \and 
          A. Noriega-Crespo
          \inst{4}
          }

   \institute{INAF, Osservatorio Astrofisico di Arcetri, Largo E. Fermi 5, 50125 Firenze, Italy\\
              \email{sepulcre@arcetri.astro.it}
              \and
              Max-Planck-Institut f\"ur Radioastronomie, Auf dem H\"ugel 69, 53121 Bonn, Germany
              \and
              Departamento de Astronom\'ia, Universidad de Chile, Casilla 36-D, Santiago, Chile
              \and
              Spitzer Science Center, California Institute of Technology, Mail Code 220-6, Pasadena, CA 91125, USA
              \and
              INAF-Istituto Fisica Spazio Interplanetario, V. Fosso del Cavaliere 100, 00133 Roma, Italy
             }

   \date{Received ; accepted}

 
  \abstract
   {Some theoretical models propose that O-B stars form via accretion, in a similar fashion to low-mass stars. Jet-driven molecular outflows play an important role in this scenario, and their study can help to understand the process of high-mass star formation and the different evolutionary phases involved.}
   {Observations towards low-mass protostars so far favour an evolutionary picture in which jets are always associated with Class 0 objects while more evolved Class I/II objects show less evidence of powerful jets. The present study aims at checking whether an analogous picture can be found in the high-mass case.}
   {The IRAM~30-m telescope (Spain) has been used to perform single-pointing SiO(2--1) and (3--2) observations towards a sample of 57 high-mass molecular clumps in different evolutionary stages. Continuum data at different wavelengths, from mid-IR to 1.2~mm, have been gathered to build the spectral energy distributions of all the clumps and estimate their bolometric luminosities.}
   {SiO emission at high velocities, characteristic of molecular jets, is detected in 88\% of our sources, a very high detection rate indicating that there is ongoing star formation activity in most of the sources of our sample. The SiO(2--1) luminosity drops with \lm, which suggests that jet activity declines as time evolves. This represents the first clear evidence of a decrease of SiO outflow luminosity with time in a homogeneous sample of high-mass molecular clumps in different evolutionary stages. The SiO(3--2) to SiO(2--1) integrated intensity ratio shows only minor changes with evolutionary state.}
   {}


   \maketitle
%

\section{Introduction}

To understand the formation mechanism of O-B stars, several models which assume high accretion rates (e.g. McKee \& Tan \cite{mckee}) and/or accretion through massive discs (e.g. Krumholz et al. \cite{krum05}) have been proposed. So far, a number of deeply embedded massive disc/outflow systems (see Cesaroni et al. \cite{cesa07}) have been found, lending support to such models. However, a detailed picture of the different evolutionary phases involved in the process is still lacking.

Jet-driven molecular outflows play an important role in the accretion scenario (Arce et al. \cite{arce}, Ray et al. \cite{ray}). An evolutionary sequence for the outflows driven by massive protostars has been proposed by Beuther \& Shepherd (\cite{beu05}), in which an initially well-collimated jet/outflow gradually evolves into a wide-angle wind as the ionising radiation powered by the central massive stellar object becomes more dominant.

L\'opez-Sepulcre et al. (\cite{yo10}; LS10 hereafter) studied a sample of high-mass molecular clumps in different evolutionary stages, in the HCO$^{+}$(1--0), HCN(1--0), and C$^{18}$O(2--1) transitions with the IRAM 30-m telescope. The detection of outflows was inferred from the presence of extended non-gaussian wings in the HCO$^{+}$(1--0) spectra, together with the spatial distribution of the blue and red lobes. The results show that molecular outflows are present in 75\% of the sources.

\vspace{-1mm}
Although HCO$^{+}$ can trace molecular outflows, contamination from the infalling envelope makes it difficult to separate the outflow component and determine the collimation degree. Further investigation in a reliable jet tracer is needed. SiO emission is ideal for this purpose, because its formation is attributed to sputtering or vaporisation of Si atoms from grains due to fast shocks (Gusdorf et al. \cite{gusdorf}, Guillet et al. \cite{guillet}), and thus suffers minimal contamination from quiescent or infalling envelopes. While in the low-mass case it has been possible to carry out high-spatial resolution observations in SiO, revealing well-collimated molecular jets (e.g. Gueth et al. \cite{gueth}), only a few SiO surveys have been made towards high-mass SFRs which include IR-dark sources, representative of the earliest evolutionary phases (e.g. Motte et al. \cite{motte}, Sakai et al. \cite{sak}). However, an interesting recent study by Jim\'enez-Serra et al. (\cite{js}) shows that the SiO emission towards the IR-dark cloud G35.39--0.33 is extended on parsec scale, suggesting widespread low-mass star formation.

Observations towards low-mass protostars so far point to an evolutionary picture in which jets are always associated with Class 0 objects (e.g. Gueth et al. \cite{gueth}, Codella et al. \cite{codella}), while Class I/II objects show less evidence of powerful jets, consistently with the progressive disappearance of the high-density clump around the young stellar object (YSO). The present study represents a first step to check whether an analogous picture can be found in the high-mass case.

With this in mind, we performed SiO(2--1) and (3--2) single-pointing observations towards the central positions of the LS10 sample, to search for evidence of emission caused by jets and check for evolutionary trends in the outflow detection rate and jet activity.


\section{The sample}

The sample under study is composed of 57 high-mass, parsec-scale molecular clumps. These include the 48 sources studied by LS10, which were selected from the millimetric surveys carried out by Rathborne et al. (\cite{rath}), Beuther et al. (\cite{beu02}), Fa\'undez et al. (\cite{faundez}) and Hill et al. (\cite{hill}); and 9 ultracompact H{\sc ii} (\uc) regions selected from the maser and/or continuum centimetric surveys performed by Hofner \& Churchwell (\cite{hc96}), Walsh et al. (\cite{walsh}), and Wood \& Churchwell (\cite{wc89}).

The sample has been sub-classified into two groups which supposedly represent two different evolutionary stages of the star formation process. These are, from less to more evolved phases: \textit{(i)} 20 infrared dark (IR-dark) clumps, i.e., not detected at 8~$\mu$m with the \textit{Midcourse Space eXperient} (MSX). Many of these are embedded 24~$\mu$m sources seen with the Multiband Imaging Photometer of Spitzer (MIPS) on the Spitzer Space Observatory. \textit{(ii)} 37 infrared loud (IR-loud) clumps, i.e., detected at 8~$\mu$m by MSX. Within the IR-loud sub-sample, 24 sources are known to contain at least one \uc\ region inside a radius of $\sim$20$''$. In addition, \uc\ regions have been detected at 3.6~cm with the VLA in 4 IR-dark clumps Battersby et al. (\cite{battersby}).

The following criteria are satisfied by the sources in our sample: (i) $\delta > -15^{\circ}$, (ii) $M_\mathrm{clump} \gtrsim 100$~M$_\odot$, to prevent contamination by low-mass star forming regions, and (iii) $d < 4.5$~kpc, to limit the spread in distances. The last criterion is satisfied by all but 5 IR-dark clumps which were found a posteriori to have distances greater than 4.5~kpc (LS10).

\section{SiO observations}

The IRAM~30-m telescope near Pico Veleta (Granada, Spain) was used on July 30-31 and August 1-2 2009 to observe our sample in the SiO(2--1) and SiO(3--2) lines, with rest frequencies at 86.85 and 130.3~GHz, respectively.

For each source, single-pointing observations in wobbler mode were made, with a wobbler throw of 120$''$ and a total integration time (ON~$+$~OFF) varying between 4 and 60 minutes, depending on the source. All the lines were covered at high spectral resolution using the autocorrelator VESPA.

The data were reduced using the programs CLASS and GREG of the GILDAS software package developed by the IRAM and the Observatoire de Grenoble. The spectra were smoothed to a resolution of 1.5~\kms\ to improve the signal to noise ratio. The typical 1$\sigma$ rms values of the spectra after smoothing are 0.009~K for the SiO(2--1) line, and 0.011~K for the SiO(3--2) line.

\section{Continuum observations}

Continuum data at different wavelengths, ranging from mid-IR to 1.2~mm, have been gathered in order to characterise the emission of the clumps from their spectral energy distributions (SEDs). Table~\ref{tcont} summarises the telescopes, instruments, wavelengths, angular resolution, and sensitivity of the data.
\begin{table}[hbt]
\centering
\caption{Continuum data}
\begin{tabular}{lccc}
\hline
Telescope - Instrument & Wavelength & Ang. Resol. & 1$\sigma$ rms\\
  & ($\mu$m) & ($''$) & (mJy~beam$^{-1}$)\\
\hline
MSX - Spirit III  & 21.3 & 20 & $\sim$150\\
{\it Spitzer} - MIPS$^\ast$ & 24 & 6 & $\sim$3\\
IRAS & 60 & 60 & $\sim$85\\
{\it Spitzer} - MIPS$^\ast$ & 70 & 18 & $\sim$40\\
IRAS & 100 & 120 & $\sim$300\\
APEX - LABOCA$^{\dag}$ & 850 & 18 & $\sim$100\\
IRAM~30-m - MAMBO$^{\ddag}$ & 1200 & 11 & $\sim$10-15 \\
SEST~15-m - SIMBA$^{\ddag}$ & 1200 & 24 & $\sim$40-150\\
\hline
\end{tabular}
\label{tcont}
\begin{itemize}
\vspace{-2mm}
\item[$^\ast$] Spitzer Galactic plane MIPSGAL survey (Carey et al. \cite{carey05}, \cite{carey09})
\item[$^\dag$] APEX Telescope Large Area Survey of the Galaxy (ATLASGAL) survey (Schuller et al. \cite{schu})
\item[$^\ddag$] Millimetre surveys cited in Sect.~2
\end{itemize}
\end{table}
\vspace{-9mm}

\section{Results and discussion}

\subsection{Spectral Energy Distributions}

Single-temperature, modified black body functions have been fitted to the SEDs of our sources to estimate their dust temperatures, masses (derived from the dust temperature and the dust continuum emission at 1.2~mm integrated over the whole clump), and bolometric luminosities (Table~\ref{tsed}). This has been possible for 47 of the observed clumps. For the remaining 10 sources, data from fewer than three of the surveys listed in Table~\ref{tcont} were available and thus the fit was not sufficiently reliable. While the single-temperature modified black body is a simplified model, it is sufficient for our purposes and it characterises the global properties of our dusty clumps.

The luminosities ($\sim$10$^3 - 10^6$~L$_\odot$) and masses ($\sim$50 to $\sim$1000~M$_\odot$) derived are typical of high-mass SFRs. The IR-dark clumps are on average colder and less luminous (by about an order of magnitude), although their masses are comparable to those of the IR-loud clumps. We note that the bolometric luminosities obtained for the sources using IRAS data represent upper limits to the actual luminosities in that there are often several sources within one IRAS beam. Similarly, upper limit luminosities were derived for 5 IR-dark sources for which only upper limits were available at 24 and/or 70~$\mu$m emission.

\subsection{SiO outflow detection rate}

SiO emission has been clearly detected above 3$\sigma$ in 88\% of the sources. We note that, even if the sources with no SiO detection correspond to spectra with $\sigma$ above the average value, the weakest detection in our sample is well above the 3$\sigma$ level of the noisiest non-detection, and therefore we do not consider our SiO detection rate to be biased by the sensitivity of the different observed spectra. Out of the 7 non-detections, 6 belong to the IR-loud group, which implies a detection rate of 84\% for the IR-loud sub-sample and 95\% for the IR-dark sub-sample. These high detection rates are similar to the one reported by Motte et al. (\cite{motte}) for a population of massive cores inside Cygnus~X.

The lines display a wide variety of intensities, widths and shapes. Figure~\ref{fspectra} presents the SiO(2--1) and (3--2) spectra of all our detected sources. In all the cases the lines have Full Widths at Zero Power (FWZP) of $\sim$10~\kms\ or more. The median FWZP of our sample is 38~\kms\ for the SiO(2--1) line, and 45~\kms\ for the SiO(3--2) line. On average, IR-dark clumps present broader SiO(2--1) wings than IR-loud clumps, with median values of 49~\kms\ for the former and 24~\kms\ for the latter. In some cases the SiO line wings extend up to velocities of about 40-50~\kms\ from the systemic velocity of the clump. Such high velocities point to the presence of molecular outflows, likely caused by jets, and therefore active star formation, in most of our sources. Table ~\ref{tpar} lists the measured SiO(2--1) and (3--2) $FWZP$s and velocity-integrated intensities for all the observed sources.

There is good agreement between the detection of SiO and the presence of \hco\ outflows as reported by LS10 (Table~\ref{tpar}): 95\% of the sources with an HCO$^{+}$ outflow present also SiO emission. There are 6 sources with clear SiO emission which have no reported HCO$^{+}$ outflow, but they all have SiO spectra with FWZP lower than about 15~\kms, and \hco\ spectra with similar widths. This is compatible with molecular outflows oriented close to the plane of the sky:  the narrowest SiO spectra likely have counterpart \hco\ spectra where the blue and red outflow wings are mixed with the low ambient velocities of the line. Our results therefore confirm that SiO is closely associated with molecular outflows. As an example, Fig.~\ref{fspt} presents the SiO(2--1) and (3--2) spectra of two of our targets (one IR-loud and one IR-dark), with an overlay of their corresponding \hco\ spectra (LS10).

\subsection{Decrease of SiO outflow activity with time}

It is reasonable to expect that the bolometric luminosity, $L_\mathrm{bol}$, of a molecular clump will increase as time evolves and the high-mass star gradually switches on, while its mass will decrease due to the effect of molecular outflows, winds, and H{\sc ii} regions (e.g. McKee \& Tan \cite{mckee02}). Therefore, the luminosity to mass ratio, \lm, can be considered to be a time estimator which, in addition, gets rid of distance effects. We will then consider \lm\ to be a measure of time or evolutionary phase, with lower values corresponding to less evolved stages of cluster formation.

\begin{figure}[hbt]
\centering
\includegraphics[scale=0.4]{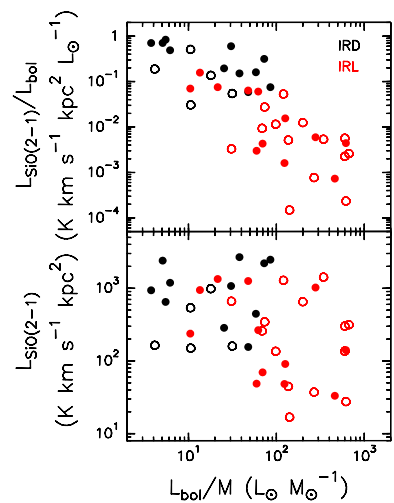}
\caption{\textit{Top}: SiO(2--1) to bolometric luminosity ratio, $L_\mathrm{SiO(2-1)}/L_\mathrm{bol}$ , against \lm. \textit{Bottom}: SiO(2--1) luminosity, $L_\mathrm{SiO(2-1)}$ , against \lm. IR-dark and IR-loud clumps are marked as black and red circles, respectively. Open circles depict sources for which the derived luminosity represents an upper limit (see Sect.~5.1).}
\label{fsio}
\end{figure}

From the integrated intensity of the SiO(2--1) line, the SiO(2--1) luminosity, $L_\mathrm{SiO(2-1)}$, has been calculated. Figure~\ref{fsio} (top) displays a plot of $L_\mathrm{SiO(2-1)}/L_\mathrm{bol}$ against \lm\ for the 47 sources where the SED could be fitted by a modified black body function (see Sect.~4.1). IR-dark and IR-loud clumps are depicted in black and red, respectively. One notes that the two types of object occupy two distinct areas of the plot. This plot shows that the most luminous SiO outflows are associated with the youngest stages of high-mass star and cluster formation. In the lower panel of Fig.~\ref{fsio}, $L_\mathrm{SiO(2-1)}$ is plotted instead of $L_\mathrm{SiO(2-1)}/L_\mathrm{bol}$. We note that the results in the upper panel of Fig.~\ref{fsio} have an inherent bias in that the bolometric luminosity appears both in the denominator of the ordinate and in the numerator of the abscissa. This however is not the case in the lower panel, where nonetheless one observes a dearth of points in the lower left hand corner. Although we do not show it here, a similar trend is found when the outflow terminal velocity is plotted against \lm.

\begin{figure}[hbt]
\centering
\includegraphics[scale=0.4]{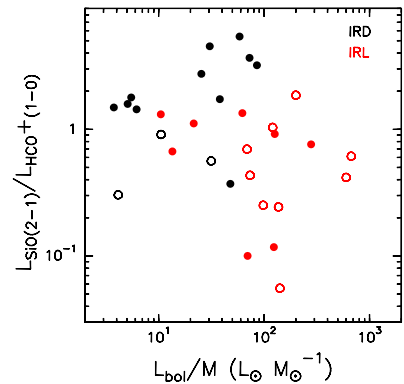}
\caption{Ratio of SiO(2--1) luminosity to \hco\ line wing luminosity, $L_\mathrm{SiO(2-1)}/L_\mathrm{HCO^{+}(1-0)}$, against \lm. The same symbols as in Fig.~\ref{fsio} apply here.}
\label{frat}
\end{figure}

The trend seen in Fig.~\ref{fsio} may be interpreted as a decrease in the SiO abundance with time (as suggested by Sakai et al. \cite{sak}), a decrease in the jet/outflow mass with time, or a combination of both. In an attempt to discriminate between these possibilities, we present in Fig.~\ref{frat} a plot of SiO(2--1) to \hco\ luminosity ratio, $L_\mathrm{SiO(2-1)}/L_\mathrm{HCO^{+}(1-0)}$, against \lm. $L_\mathrm{HCO^{+}(1-0)}$ corresponds to the emission under the line wings defined by LS10. It can be seen that IR-dark sources lie higher in the plot, while high \lm\ sources are on average lower and have greater dispersion in $L_\mathrm{SiO(2-1)}/L_\mathrm{HCO^{+}(1-0)}$. This difference between less evolved and more evolved objects favours the interpretation in which SiO abundance decreases with time, since otherwise one would expect all the points in Fig.~\ref{frat} to cluster around a constant value of $L_\mathrm{SiO(2-1)}/L_\mathrm{HCO^{+}(1-0)}$. Moreover, LS10 reported similar outflow masses among the IR-dark and the IR-loud sub-samples, and therefore there is no significant difference in the outflow mass between more evolved and less evolved clumps. Given that SiO is thought to be produced in fast shocks which can be effectively generated by fast collimated jets, this result suggests that the earliest phases of star formation are dominated by well-collimated jets, which gradually de-collimate and lose power. Such a decrease in the SiO outflow luminosity with time seems analogous to what is found for low-mass SFRs (e.g. Gibb et al. \cite{gibb04}), where the most intense SiO emission is detected towards the less evolved Class 0 objects. This similarity between the low-mass and the high-mass regime lends support to the disc accretion models as the mechanism to form high-mass stars.

It could be argued that the SiO emission arising from the high \lm\ sources comes from low-mass YSOs within the region and not from their high-mass neighbours. However, LS10 reported outflow masses and momentum rates consistent with high-mass stars, independently of the evolutionary phase. Therefore, the correlation seen in Fig.~\ref{fsio} most likely represents an evolutionary line for molecular outflows driven by massive YSOs, compatible with the scenario proposed by Beuther \& Shepherd (\cite{beu05}) in which initially well collimated outflows gradually disappear to make way for a wide-angle wind-dominated outflow. Our result is consistent with the findings of Miettinen et al. (\cite{fin}), who measured a decrease of the SiO abundance with core temperature in a sample of 15 \uc\ regions and concluded that SiO outflows are less frequent in more evolved, warmer cores. Given that our sample, which includes IR-dark sources, presumably covers a wider range of evolutionary phases, this represents the first time that clear evidence of a decrease in jet/outflow activity with time is found in a sample of high-mass SFRs.

\begin{figure}[htb]
\centering
\includegraphics[scale=0.4]{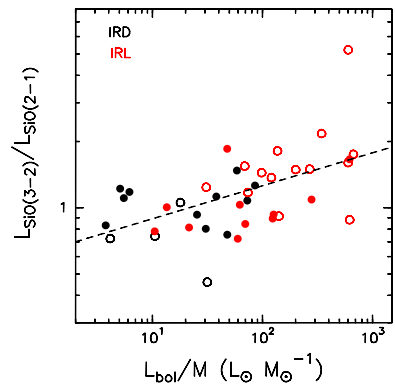}
\caption{SiO(3--2) to SiO(2--1) luminosity ratio, $L_\mathrm{SiO(3-2)}/L_\mathrm{SiO(2-1)}$, against \lm. The dashed line represents the best fit to the points (see text). The same symbols as in Fig.~\ref{fsio} apply here.}
\label{fexc}
\end{figure}

A final remark concerns the excitation conditions of the jet environment. We checked for a possible dependence of the SiO(3--2) to SiO(2--1) intensity ratio on velocity, but a well defined trend has not been discerned. Further, we present in Fig.~\ref{fexc} a plot of SiO(3--2) to SiO(2--1) luminosity ratio, $L_\mathrm{SiO(3-2)}/L_\mathrm{SiO(2-1)}$, as a function of \lm. Such a ratio changes by at most a factor 4 over the sampled range of \lm. The points follow a weak trend which can be described by the function $L_\mathrm{SiO(3-2)}/L_\mathrm{SiO(2-1)} = 0.63 ($\lm$)^{0.15}$, with a correlation coefficient of 0.38, obtained from a least squares fit to all the points. Notice, however, that the the best fit results depend on whether the open circles (representing upper limits to the bolometric luminosity) are included or not, and therefore the correlation found should be confirmed using a more statistically reliable sample. The trend, if real, suggests that the excitation temperature of the post-shock environment increases only marginally with time, which may be explained by a small increase in density with time. To this end, we used the non-LTE excitation code RADEX with an escape probability formalism for the radiative transfer (Van der Tak et al. \cite{radex}) coupled with the LAMDA database (Sch\"oier et al. \cite{lamda}). Assuming that (i) both SiO lines trace the same region, and (ii) the beam filling factor equals unity, then, for a kinetic temperature between 50 and 150~K, we infer densities between $5 \times 10^4$ and 10$^6$~cm$^{-3}$. The total SiO column densities lie in the range 10$^{12}$ - 10$^{13}$~cm$^{-2}$. A more exhaustive study of the level of excitation of SiO would require high angular resolution mapping of higher $J$ transitions.

\section{Conclusions}

We have observed SiO(2--1) and SiO(3--2) with the IRAM~30-m telescope towards a sample of 57 high-mass molecular clumps in different evolutionary stages, and characterised their IR and (sub)mm emission using continuum data from the archive and a modified grey body function model to fit their SEDs. High-velocity SiO emission has been detected in 88\% of our sample at high velocities, implying star formation is actively taking place in most of our targets. A tight inverse correlation is found between $L_\mathrm{SiO(2-1)}/L_\mathrm{bol}$ and \lm, which is interpreted as the presence of an evolutionary line in which jet activity fades as time evolves. This picture is analogous to what is found in the low-mass regime, which suggests that high-mass stars may form generally via disc accretion as low-mass stars. Our result is consistent with the evolutionary picture suggested by Beuther \& Shepherd (\cite{beu05}) for massive molecular outflows. For the first time, evidence of a clear decrease of SiO outflow luminosity with time has been measured in a sample of high-mass star forming clumps in different evolutionary stages, covering both IR-dark and IR-loud clumps. A relatively constant luminosity ratio of SiO(3--2) to (2--1) integrated intensity is found in our sample, suggesting that the post-shock density does not vary greatly. We note, however, that high-angular resolution imaging is necessary to confirm our conclusions.

\begin{acknowledgements}
A.L.S. acknowledges support from the FP6 Marie-Curie Research Training Network ``Constellation: the origin of stellar masses'' (MRTN-CT-2006-035890). L.B. acknowledges support from FONDAP Center for Astrophysics 15010003. We are grateful to the staff of IRAM-Granada for the help provided during the observations at the 30-m telescope. We also thank our referee for his valuable comments and suggestions.
\end{acknowledgements}

\Online

\begin{appendix}

\section{Tables}

\begin{table}[!ht]
\centering
\caption{Properties of the clumps derived from modified black body fits to their SEDs}
\begin{tabular}{lccc}
\hline
\hline
Source & $T_\mathrm{dust}$ & $M$ & $L_\mathrm{bol}$\\
 &  (K) & (M$_\odot$) & (L$_\odot$)\\
\hline
05358+3543 & 76 & 70 & 4370\\ 
G213.61-12.6$^\ast$ & 39 & 800 & 113000\\ 
G189.78+0.34$^\ast$ & 42 & 63 & 8630\\ 
G192.60--0.05$^\ast$ & 72 & 100 & 60200\\ 
18151-1208\_1$^\ast$ & 75 & 120 & 11800\\ 
G18.15--0.28 & 105 & 43 & 33400\\ 
G18.18--0.30 & 32 & 55 & 2640\\ 
18223--1243 & 73 & 230 & 16100\\ 
18228--1312 & 95 & 240 & 29800\\ 
G19.27+0.1M2 & 29 & 57 & 1450\\ 
G19.27+0.1M1 & 33 & 47 & 2760\\ 
18232--1154 & 53 & 61 & 19800\\ 
18236--1205 & 60 & 320 & 3370\\ 
G19.61--0.24A$^\ast$ & 59 & 765 & 264000\\ 
G20.08--0.1 & 66 & 410 & 19700\\ 
18264--1152 & 59 & 440 & 5970\\ 
G23.60+0.0M1 & 21.5 & 660 & 3370\\ 
18316--0602 & 48 & 610 & 171000\\ 
G23.60+0.0M2 & 26 & 58 & 1780\\ 
18317--0513 & 79 & 270 & 16200\\ 
G24.08+0.0M2$^\ast$ & 31 & 93 & 2930\\ 
G24.33+0.1M1 & 56 & 460 & 17600\\ 
G24.33+0.1M4$^\ast$ & 41 & 460 & 4890\\ 
G24.33+0.1M2$^\ast$ & 28 & 400 & 7190\\ 
G24.60+0.1M2 & 21 & 390 & 2420\\ 
G24.60+0.1M1 & 22 & 140 & 772\\ 
G25.04--0.2M1 & 20 & 350 & 1320\\ 
G25.04--0.2M4$^\ast$ & 33 & 210 & 869\\ 
G25.04--0.2M2$^\ast$ & 25 & 100 & 1060\\ 
G28.28--0.35 & 109 & 97 & 45100\\ 
G34.43+0.2M1 & 42 & 380 & 32700\\ 
18507+0121 & 61 & 820 & 17700\\ 
G34.43+0.2M3 & 35 & 95 & 6940\\
G34.24+0.13 & 48 & 51 & 31700\\ 
18517+0437$^\ast$ & 43 & 400 & 27600\\ 
19035+0641$^\ast$ & 74 & 46 & 5830\\ 
19095+0930 & 61 & 260 & 52100\\ 
G43.89--0.38$^\ast$ & 57 & 88 & 53100\\ 
G61.48+0.09A$^\ast$ & 66 & 189 & 117000\\ 
20216+4107$^\ast$ & 72 & 17 & 3180\\ 
20332+4124 & 79 & 180 & 23900\\ 
22134+5834$^\ast$ & 82 & 55 & 12600\\ 
22570+5912\_1$^\ast$ & 82 & 180 & 48400\\ 
23033+5951 & 71$^\ast$ & 170 & 12500\\ 
NGC7538B$^\ast$ & 37 & 6480 & 200000\\ 
23139+5939$^\ast$ & 71 & 200 & 24100\\
23151+5912$^\ast$ & 98 & 180 & 121000\\ 
\hline
\end{tabular}
\begin{itemize}
\item[$^\ast$] Sources whose derived luminosities are upper limits (see Sect.~5.1)
\end{itemize}
\label{tsed}
\end{table}

\begin{table*}[!ht]
\centering
\caption{Parameters of the SiO(2--1) and (3--2) lines, and HCO$^{+}$ outflow detection}
\begin{tabular}{llcccccccc}
\hline
\hline
Sub-class & Source & RA & Dec & $V_\mathrm{LSR}$$^\ast$ & \multicolumn{2}{c}{SiO(2--1)} & \multicolumn{2}{c}{SiO(3--2)} & HCO$^{+}$\\
 &  &  &  &  & $FWZP$$^\dag$ & $\int F_\mathrm{v} dv$$^\dag$ & $FWZP$$^\dag$ & $\int F_\mathrm{v} dv$$^\dag$ & outflow?$^\ddag$\\
 &  & (J2000.0) & (J2000.0) & (km~s$^{-1}$) & (km~s$^{-1}$) & (K~km~s$^{-1}$) & (km~s$^{-1}$) & (K~km~s$^{-1}$) & (Y/N)\\
\hline
IRL & 05358+3543 & 05:39:12.20 & +35:45:52.0 & $-$15.8 & 38.3 & 6.5 & 50.2 & 6.8 & Y\\
IRL & G213.61-12.6 & 06:07:49.23 & $-$06:22:40.6 & 10.6 & 9.2 & 0.3 & 5.8 & 0.3 & Y\\
IRL & G189.78+0.34 & 06:08:35.35 & +20:39:04.3 & 9.2 & 15.2 & 1.1 & 21.0 & 2.0 & Y\\
IRL & G192.58-0.04 & 06:12:52.90 & +18:00:35.0 & 9.1 & 35.9 & 1.7 & 17.0 & 1.4 & Y\\
IRL & G192.60-0.05 & 06:12:54.00 & +17:59:23.0 & 7.4 & 29.2 & 1.6 & 23.1 & 2.6 & Y\\
IRD & 18151-1208\_2 & 18:17:50.50 & $-$12:07:55.0 & 29.8 & 84.3 & 5.7 & 103.1 & 11.6 & Y\\
IRL & 18151-1208\_3 & 18:17:52.30 & $-$12:06:56.0 & 30.7 & --- & --- & --- & --- & N\\
IRL & 18151-1208\_1 & 18:17:58.00 & $-$12:07:27.0 & 33.3 & 36.3 & 1.2 & 31.1 & 1.7 & Y\\
IRL & G18.15-0.28 & 18:25:02.41 & $-$13:15:25.7 & 55.8 & --- & --- & --- & --- & N\\
IRD & G18.18-0.30 & 18:25:07.30 & $-$13:14:23.0 & 50.0 & 35.9 & 1.8 & 29.8 & 1.4 & Y\\
IRL & 18223-1243 & 18:25:10.87 & $-$12:42:26.8 & 45.2 & 21.2 & 0.4 & 13.0 & 0.3 & Y\\
IRL & 18228-1312 & 18:25:42.35 & $-$13:10:18.1 & 33.1 & 15.2 & 0.4 & 27.9 & 0.4 & Y\\
IRD & G19.27+0.1M2 & 18:25:52.60 & $-$12:04:48.0 & 26.9 & 66.3 & 4.0 & 51.0 & 3.7 & Y\\
IRD & G19.27+0.1M1 & 18:25:58.50 & $-$12:03:59.0 & 26.5 & 75.9 & 6.2 & 69.9 & 9.1 & Y\\
IRL & 18232-1154 & 18:26:04.37 & $-$11:52:33:4 & 24.4 & --- & --- & --- & --- & ---\\
IRL & 18236-1205 & 18:26:25.41 & $-$12:03:51.4 & 26.5 & 51.1 & 3.0 & 34.8 & 2.4 & Y\\
IRL & G19.61-0.24A & 18:27:38:16 & $-$11:56:40.2 & 42.4 & 51.1 & 7.8 & 59.0 & 17.0 & --- \\
IRL & G20.08-0.14 & 18:28:10:28 & $-$11:28:48.7 & 42.5 & 41.9 & 8.7 & 50.0 & 16.2 & ---\\
IRL & 18264-1152 & 18:29:14.40 & $-$11:50:21.3 & 43.9 & 71.1 & 6.2 & 40.1 & 6.2 & Y\\
IRD & G23.60+0.0M1 & 18:34:11.60 & $-$08:19:06.0 & 106.5 & 73.1 & 5.0 & 82.9 & 6.1 & Y\\
IRL & 18316-0602 & 18:34:20.46 & $-$05:59:30.4 & 42.5 & 75.1 & 8.5 & 60.3 & 9.3 & Y\\
IRD & G23.60+0.0M2 & 18:34:21.10 & $-$08:18:07.0 & 53.6 & 85.1 & 5.6 & 57.9 & 4.5 & Y\\
IRL & 18317-0513 & 18:34:25.94 & $-$05:10:48.6 & 42.0 & 16.4 & 0.4 & 9.8 & 0.3 & N\\
IRD & G24.08+0.0M2 & 18:34:51.10 & $-$07:45:32.0 & 52.0 & 21.2 & 0.9 & 17.0 & 0.4 & Y\\
IRD & G24.33+0.1M1 & 18:35:07.90 & $-$07:35:04.0 & 113.6 & 49.1 & 4.8 & 55.0 & 5.4 & Y\\
IRD & G24.33+0.1M4 & 18:35:18.56 & $-$07:37:26.2 & 114.3 & 8.4 & 0.3 & --- & --- & N\\
IRD & G24.33+0.1M2 & 18:35:34.50 & $-$07:37:28.0 & 118.6 & 33.1 & 1.6 & 44.6 & 1.7 & N\\
IRD & G24.60+0.1M2 & 18:35:35.70 & $-$07:18:09.0 & 115.3 & 53.1 & 2.1 & 43.0 & 2.4 & Y\\
IRD & G24.60+0.1M1 & 18:35:40.20 & $-$07:18:37.0 & 53.2 & 49.9 & 3.8 & 67.0 & 4.2 & Y\\
IRD & G25.04-0.2M1 & 18:38:10.20 & $-$07:02:34.0 & 63.8 & 66.3 & 4.0 & 52.1 & 3.4 & Y\\
IRD & G25.04-0.2M4 & 18:38:13.70 & $-$07:03:12.0 & 63.8 & 18.4 & 0.7 & 13.0 & 0.5 & Y\\
IRD & G25.04-0.2M2 & 18:38:17.70 & $-$07:02:51.0 & 63.9 & 30.0 & 2.3 & 24.2 & 1.7 & Y\\
IRL & G28.28-0.35 & 18:44:14.20 & $-$04:17:59.0 & 48.7 & 8.4 & 0.3 & --- & --- & N\\
IRD & G34.43+0.2M1 & 18:53:18.00 & +01:25:23.0 & 58.1 & 75.1 & 14.4 & 63.0 & 18.2 & Y\\
IRL & G34.26+0.15 & 18:53:18.44 & +01:14:57.8 & 58.4 & 25.2 & 6.6 & 24.2 & 13.4 & ---\\
IRL & 18507+0121 & 18:53:19.58 & +01:24:37.1 & 58.2 & 57.1 & 7.8 & 52.1 &6.4  & Y\\
IRD & G34.43+0.2M3 & 18:53:20.40 & +01:28:23.0 & 59.4 & 85.9 & 12.9 & 72.0 & 13.9 & Y\\
IRL & G34.24+0.13 & 18:53:21.70 & +01:13:37.0 & 57.2 & 18.0 & 0.9 & 34.0 & 1.5 & N\\
IRL & 18517+0437 & 18:54:14.30 & +04:41:40.0 & 44.1 & 13.2 & 2.4 & 26.0 & 3.8 & Y\\
IRD & G34.77-0.6M2 & 18:56:49.41 & +01:23:16.0 & 42.2 & --- & --- & --- & --- & N\\
IRD & G35.39-0.3M4 & 18:57:07.90 & +02:08:23.0 & 44.6 & 15.2 & 0.5 & 9.0 & 0.5 & N\\
IRL & G35.20-1.74 & 19:01:46.40 & +01:13:24.5 & 45.0 & 33.1 & 0.7 & 32.2 & 1.2 & ---\\
IRD & G38.95-0.5M1 & 19:04:07.40 & +05:08:48.0 & 42.3 & 32.4 & 1.8 & 14.9 & 1.1 & Y\\
IRL & 19035+0641 & 19:06:01.60 & +06:46:43.0 & 32.8 & 24.4 & 1.5 & 30.0 & 1.4 & Y\\
IRL & 19095+0930 & 19:11:54.02 & +09:35:52.0 & 43.9 & 42.3 & 4.8 & 55.8 & 7.1 & Y\\
IRL & G43.89-0.38 & 19:14:26.07 & +09:22:33.9 & 53.7 & 41.9 & 1.4 & 47.8 & 7.1 & ---\\
IRL & G61.48+0.09A & 19:46:49.11 & +25:12:48.0 & 21.5 & 11.2 & 0.5 & 12.0 & 0.5 & ---\\
IRL & G75.78-0.34 & 20:21:44.01 & +37:26:39.4 & 0.3 & 36.3 & 4.4 & 40.9 & 6.9 & ---\\
IRL & 20216+4107 & 20:23:23.60 & +41:17:38.0 & $-$1.6 & --- & --- & --- & --- & Y\\
IRL & 20332+4124 & 20:34:59.90 & +41:34:49.0 & $-$2.7 & --- & --- & --- & --- & N\\
IRL & 22134+5834 & 22:15:09.30 & +58:49:06.0 & $-$18.3 & --- & --- & --- & --- & Y\\
IRL & 22570+5912\_2 & 22:58:58.75 & +59:27:29.2 & $-$47.8 & 12.0 & 0.5 & 22.1 & 0.7 & N\\
IRL & 22570+5912\_1 & 22:59:05.90 & +59:28:19.0 & $-$45.6 & --- & --- & 5.0 & 0.2 & N\\
IRL & 23033+5951 & 23:05:25.50 & +60:08:06.0 & $-$53.1 & 33.9 & 2.2 & 52.9 & 2.6 & Y\\
IRL & NGC7538B & 23:13:45.25 & +61:28:10.2 & $-$57.0 & 38.3 & 4.3 & 57.1 & 5.3 & ---\\
IRL & 23139+5939 & 23:16:11.12 & +59:55:30.8 & $-$44.1 & 61.1 & 4.4 & 59.0 & 6.1 & Y\\
IRL & 23151+5912 & 23:17:21.00 & +59:28:49.0 & $-$54.5 & 25.2 & 0.8 & 27.1 & 1.3 & Y\\
\hline
\end{tabular}
\begin{itemize}
\item[$^\ast$] Systemic velocity of the clump as measured from the peak velocity of the optically thin H$^{13}$CO$^{+}$(1--0) line, observed simultaneously to SiO(2--1) and SiO(3--2)
\item[$^\dag$] --- : not detected
\item[$^\ddag$] --- : no observation made towards this source (see Sect.~2)
\end{itemize}
\label{tpar}
\end{table*}

\section{Figures}

\begin{figure}[!hbt]
\centering
\includegraphics[scale=0.65]{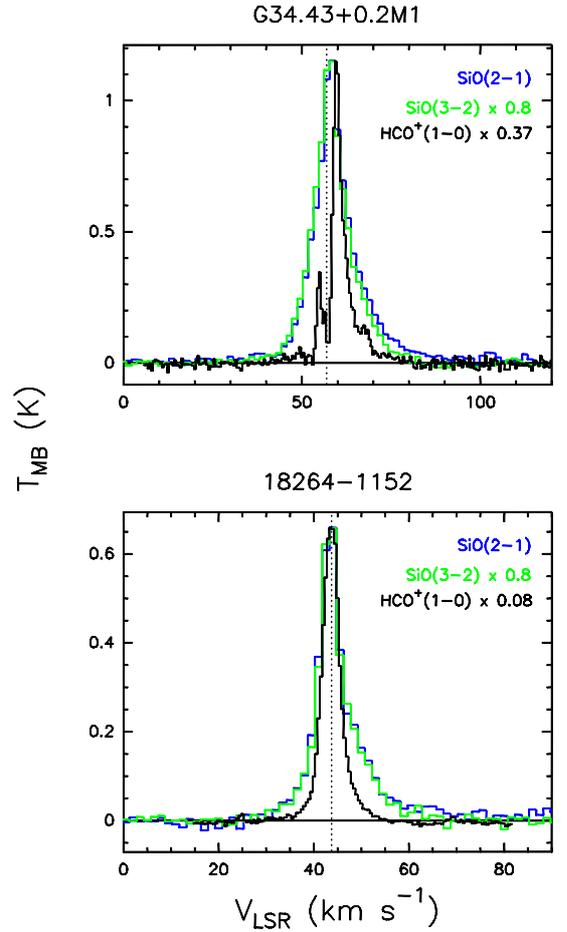}
\caption{SiO(2--1) ({\it blue}), SiO(3--2) ({\it green}), and \hco ({\it black})  spectra observed towards the IR-dark source G34.43+0.2M1 and the IR-loud source 18264--1152. The vertical dotted lines mark the systemic velocity of the clump, listed in col.~5 of Table~\ref{tpar}.}
\label{fspt}
\end{figure}

\begin{figure*}[htb]
\centering
\includegraphics[scale=1]{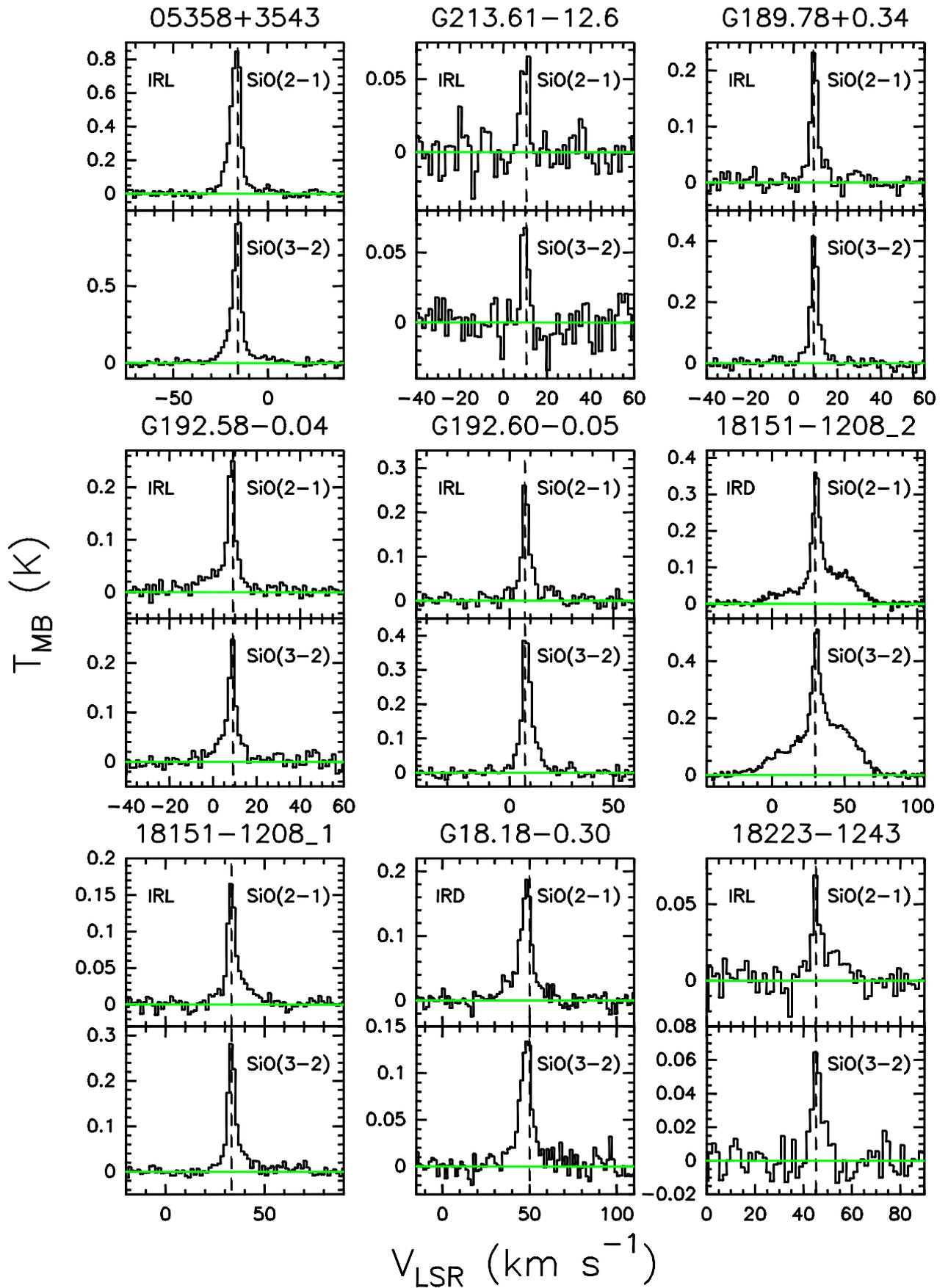}
\caption{SiO(2--1) and (3--2) spectra. The dashed vertical line marks the systemic velocity of the clump, listed in col.~5 of Table~\ref{tpar}.}
\label{fspectra}
\end{figure*}

\addtocounter{figure}{-1}
\begin{figure*}[htb]
\centering
\includegraphics[scale=1]{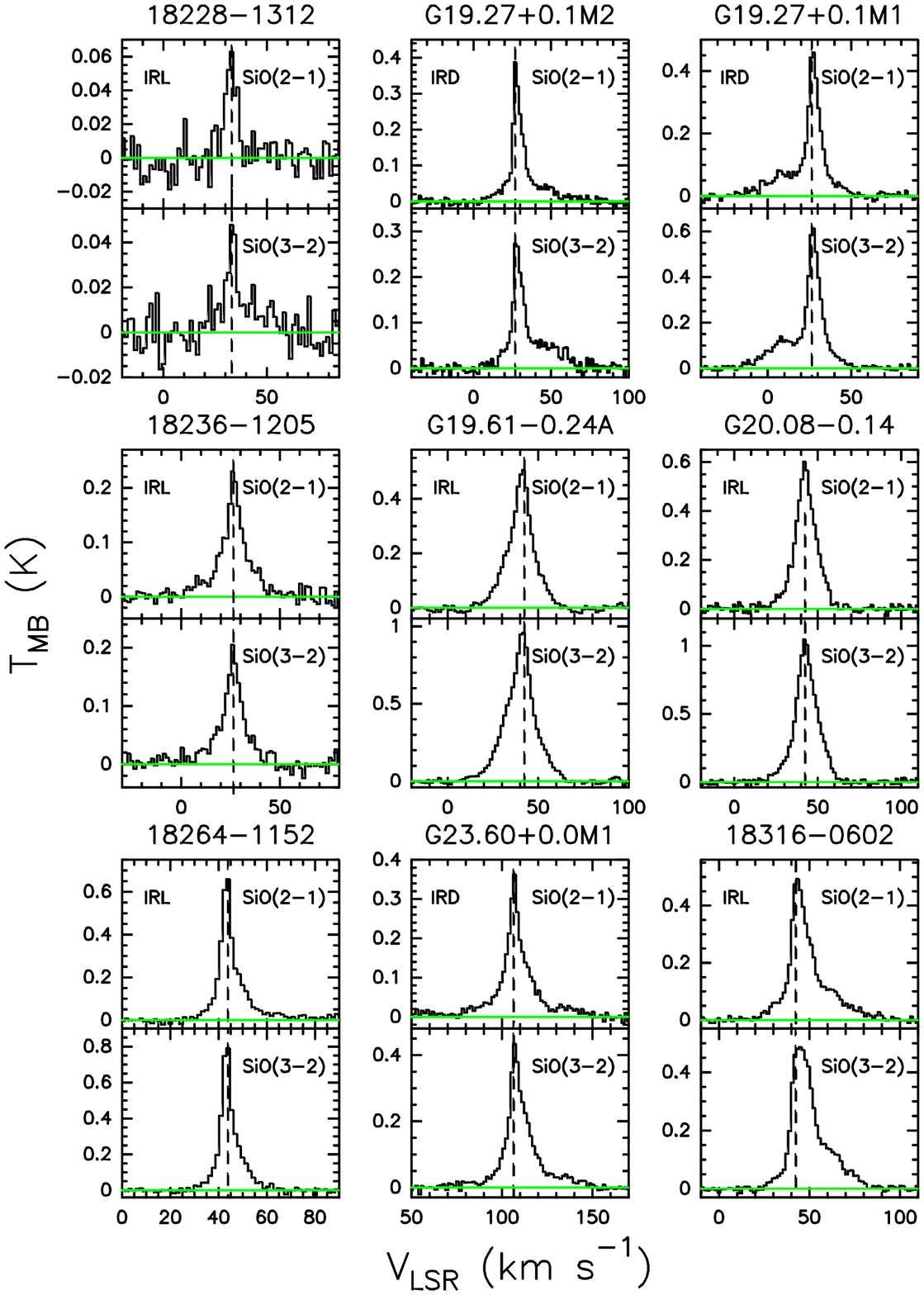}
\caption{\textit{Continued.}}
\end{figure*}

\addtocounter{figure}{-1}
\begin{figure*}[htb]
\centering
\includegraphics[scale=1]{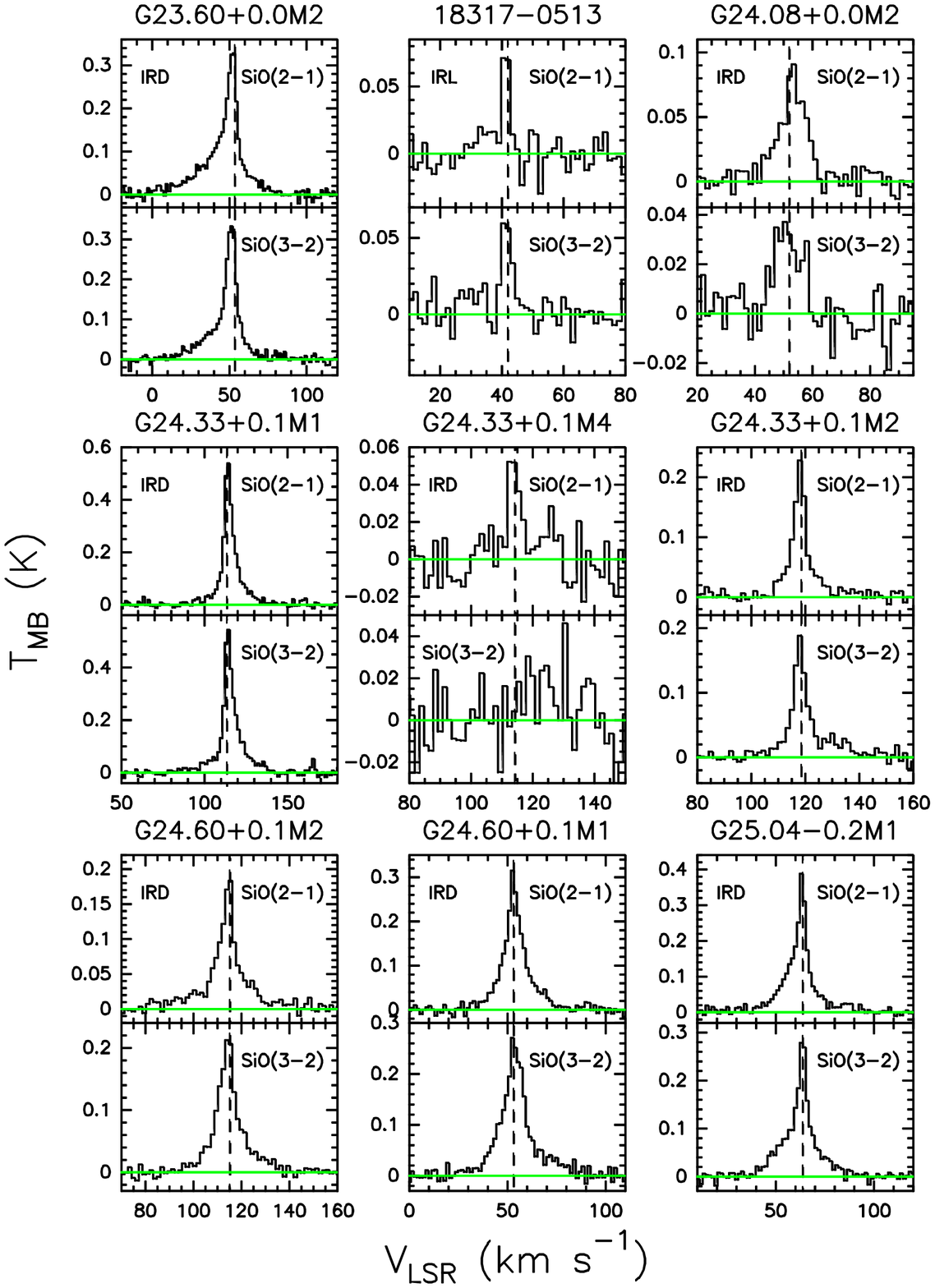}
\caption{\textit{Continued.}}
\end{figure*}

\addtocounter{figure}{-1}
\begin{figure*}[htb]
\centering
\includegraphics[scale=1]{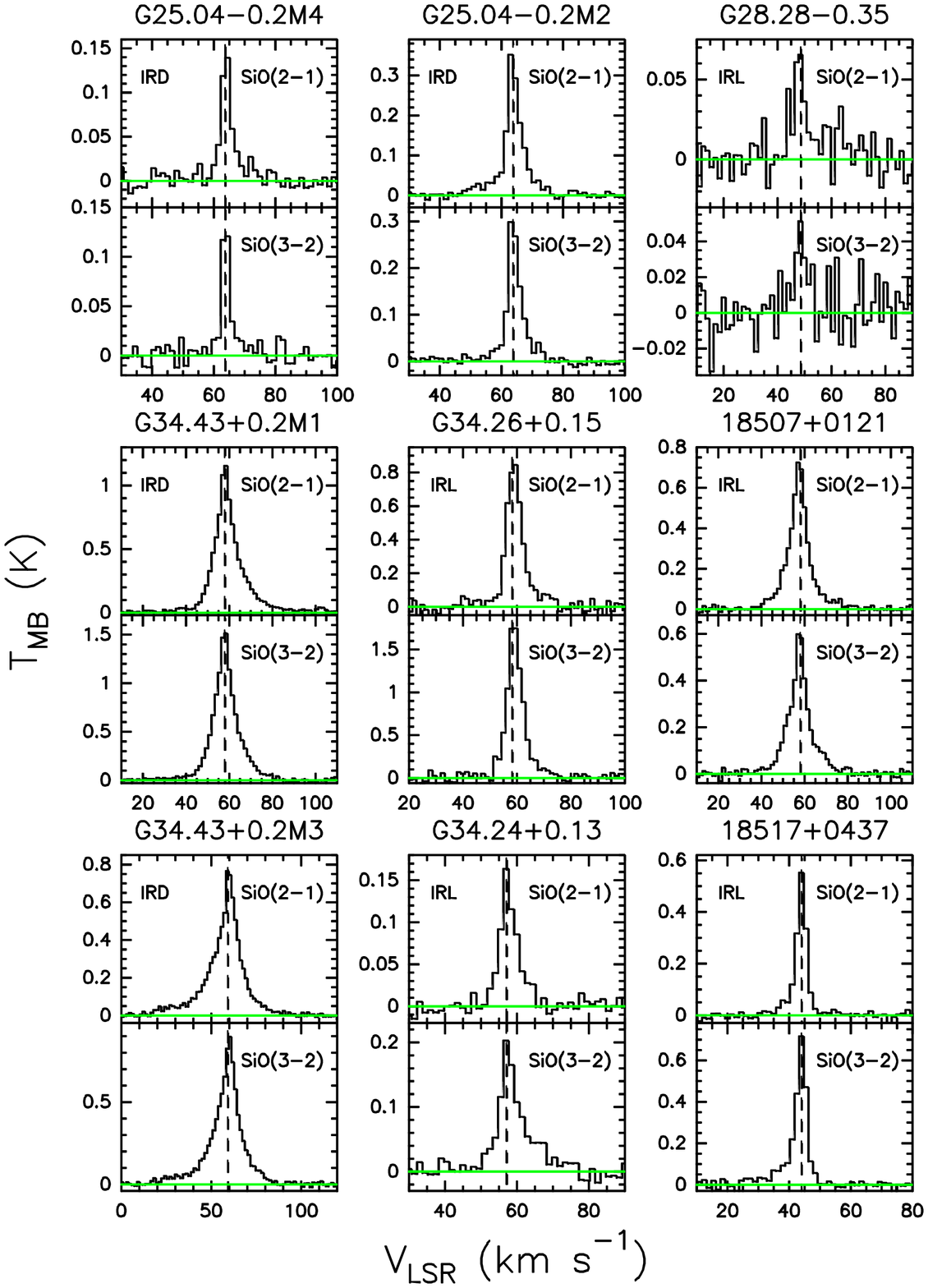}
\caption{\textit{Continued.}}
\end{figure*}

\addtocounter{figure}{-1}
\begin{figure*}[htb]
\centering
\includegraphics[scale=1]{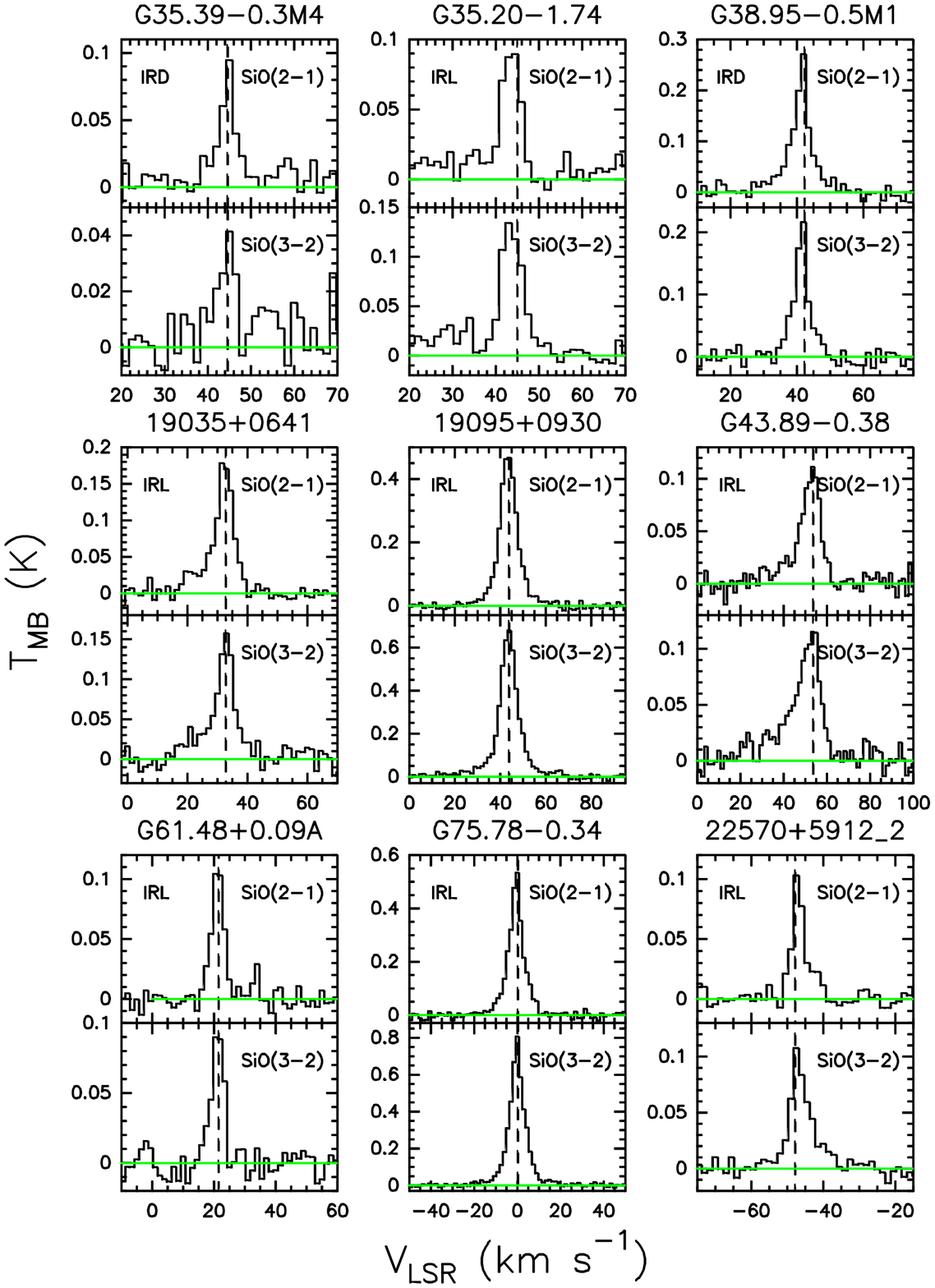}
\caption{\textit{Continued.}}
\end{figure*}

\addtocounter{figure}{-1}
\begin{figure*}[htb]
\includegraphics[scale=1]{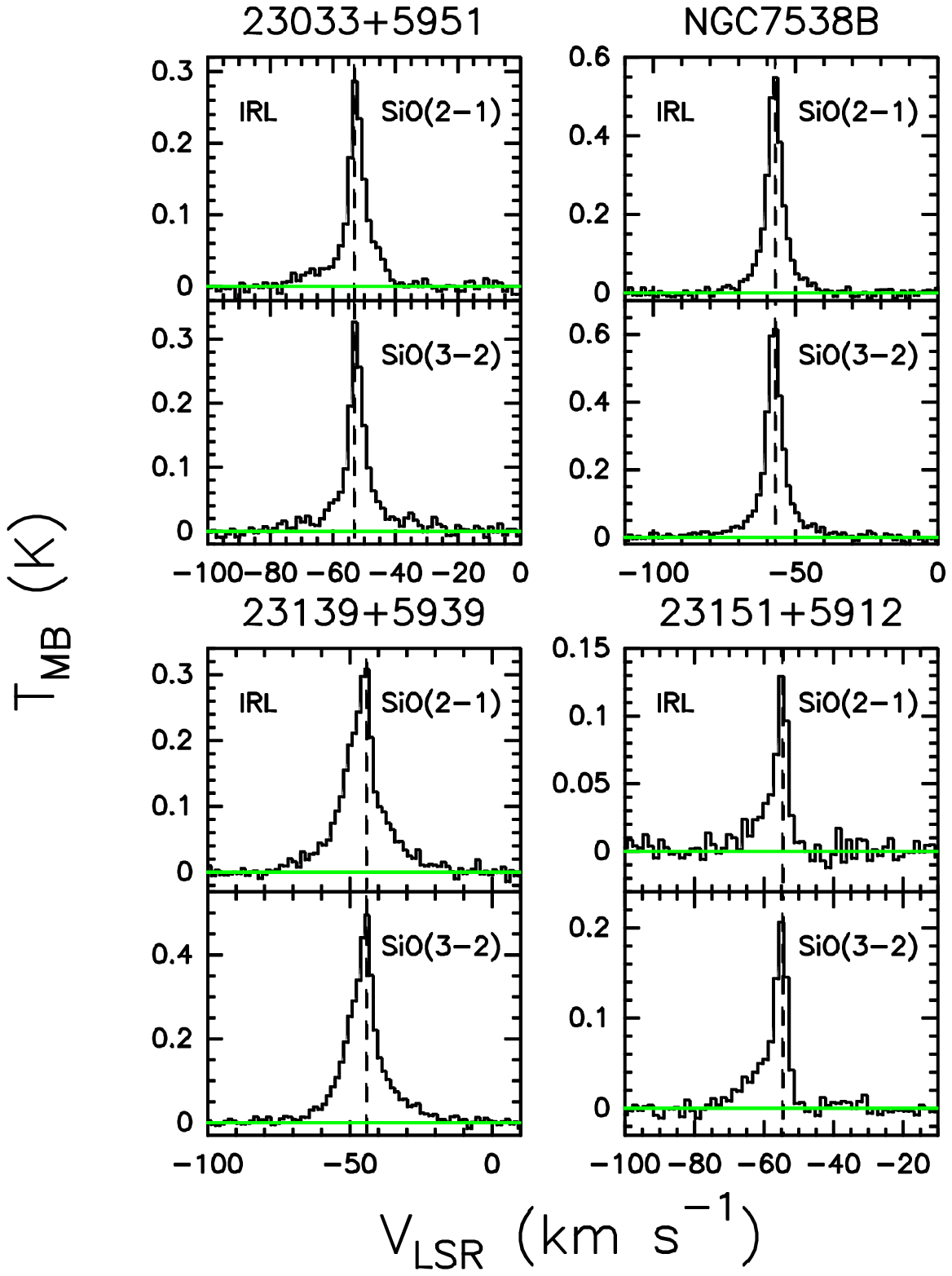}
\caption{\textit{Continued.}}
\end{figure*}

\end{appendix}


\begin{thebibliography}{}
\bibitem[2007]{arce} Arce, H.G., Shepherd, D., Gueth, F. et al. 2007, PPV, p. 245
\bibitem[2010]{battersby} Battersby, C., Bally, J., Jackson, J.M., et al. 2010, arXiv1008.0871B
\bibitem[2007]{beusrid} Beuther, H. \& Sridharan, T.K. 2007, ApJ 668, 348
\bibitem[2002]{beu02} Beuther, H., Schilke, P., Menten, K.M. et al. 2002, ApJ 566, 945
\bibitem[2005]{beu05} Beuther \& H, Shepherd, D. 2005, CCSF, p.~105
\bibitem[2009]{carey09} Carey, S.J., Noriega-Crespo, A., Mizuno, D.R., et al. 2009, PASP 121, 76
\bibitem[2005]{carey05} Carey, S.J., Noriega-Crespo, A., Price, S., et al. 2005, BAAS, 37, 1252
\bibitem[2007]{cesa07} Cesaroni, R., Galli, D., Lodato, G., Walmsley, C.M., \& Zhang, Q. 2007, in {\it Protostars and Planets V}, 197, Univ. Arizona Press, Tucson
\bibitem[2007]{codella} Codella, C., Cabrit, S., Gueth, F. et al. 2007, A\&A 462, L53
\bibitem[2004]{faundez} Fa\'undez, S., Bronfman, L., Garay, G. et al. 2004, A\&A 426, 97
\bibitem[2007]{gibb} Gibb, A.G., Davis, C.J., \& Moore, T.J.T. 2007, MNRAS 382, 1213
\bibitem[2004]{gibb04} Gibb, A.G., Richer, J.S., Chandler, C.J., \& Davis, C.J. 2004, ApJ 603, 198
\bibitem[2005]{giveon} Giveon, U., Becker, R.H., Helfand, D.J., \& White, R.L. 2005, ApJ 129, 348
\bibitem[1999]{gueth} Gueth, F. \& Guilloteau, S. 1999, A\&A 343, 571
\bibitem[2009]{guillet} Guillet, V., Jones, A.P., Pineau Des For\^ets, G. 2009, A\&A 497, 145
\bibitem[2008]{gusdorf} Gusdorf, A., Cabrit, S., Flower, D.R. et al. 2008, A\&A 482, 809
\bibitem[2005]{hill} Hill, T., Burton, M.G., Minier, V. et al. 2005, MNRAS 363, 405
\bibitem[1996]{hc96} Hofner, P. \& Churchwell, E. 1996, A\&AS 120, 283
\bibitem[2010]{js} Jim\'enez-Serra, I., Caselli, P., Tan, J.C., et al. 2010, MNRAS 406, 187
\bibitem[2005]{krum05} Krumholz, M.R., McKee, C.F., \& Klein, R.I. 2005, ApJ 618, L33
\bibitem[2010]{yo10} L\'opez-Sepulcre, Cesaroni, \& R., Walmsley, M. 2010, A\&A, 517, 66 (LS10)
\bibitem[2002]{mckee02} McKee, C.F. \& Tan, J.C. 2002, Nature 416, 59
\bibitem[2003]{mckee} McKee, C.F. \& Tan, J.C. 2003, ApJ 585, 850
\bibitem[2006]{fin} Miettinen, O., Harju, J., Haikala, L.K., \& Pomr\'en, C. A\&A, 460, 721
\bibitem[2007]{motte} Motte, F., Bontemps, S., Schilke, P. et al. 2007, A\&A 476, 1243
\bibitem[2006]{rath} Rathborne, J.M., Jackson, J.M., \& Simon, R. 2006, ApJ 641, 389
\bibitem[2010]{rath10} Rathborne, J.M., Jackson, J.M., Chambers, E.T. et al. 2010, ApJ 715, 310
\bibitem[2007]{ray} Ray, T., Dougados, C., Bacciotti, F. et al. 2007, PPV, p.~231
\bibitem[2010]{sak} Sakai, T., Sakai, N., Hirota, T., \& Yamamoto, S. 2010, ApJ, 714, 1658
\bibitem[2005]{lamda} Sch\"oier F.L., van der Tak FFS., van Dishoeck E.F., \& Black J.H. 2005, A\&A, 432, 369
\bibitem[2009]{schu} Schuller, F., Menten, K.M., Contreras, Y. et al. 2009, A\&A, 504, 415
\bibitem[2007]{radex} Van der Tak F.F.S., Black J.H., Sch\"oier F.L., Jansen D.J., \& van Dishoeck, E.F. 2007, A\&A, 468, 627
\bibitem[1998]{walsh} Walsh, A.J., Burton, M.G., Hyland, A.R. et al. 1998, MNRAS 301, 640
\bibitem[1989]{wc89} Wood, D.O.S. \& Churchwell, E. 1989, ApJ 69, 831
\end{thebibliography}
\end{document}